\documentstyle[aps,twocolumn,epsfig]{revtex}
\begin{document}
\title{  Can $J/\psi$ suppression and $p_T$ broadening signal the
deconfinement transition at RHIC?}
%Revised version put into the archive
\author{\bf A. K. Chaudhuri\cite{byline}}
\address{ Variable Energy Cyclotron Centre\\
1/AF,Bidhan Nagar, Calcutta - 700 064\\}
\maketitle
\begin{abstract}

We  have analyzed the latest NA50 data on $J/\psi$ suppression in
Pb+Pb collisions at CERN SPS.  It  is  shown  that  a  QCD  based
nuclear  absorption  model,  where  $J/\psi$'s  are  absorbed  in
nuclear  medium  could  explain  the  latest  NA50  data  on  the
centrality  dependence  of  the $J/\psi$ over Drell-Yan ratio. The model
also explains the NA50 data on $J/\psi$ over minimum  bias  ratio
and  the  $p_T$  broadening  of $J/\psi$'s. A QGP based threshold
model where all the $J/\psi$'s are suppressed above  a  threshold
density,  also  explains  the  data  sets  with smeared threshold
density. Even at RHIC energy, centrality dependence  of  $J/\psi$
suppression or $p_T$ broadening could not distinguish between the
two models.
\end{abstract}


\section{Introduction}

$J/\psi$  suppression in heavy ion collisions is recognized as an
important   signal   of   the   confinement-deconfinement   phase
transition.   NA50   collaboration,   at  CERN,  is  a  dedicated
experiment,  measuring   $J/\psi$   cross   sections   in   pA/AA
collisions. Recently, they have published the results of 2000 run
of  158  AGeV  Pb+Pb  collisions  \cite{na50a1}.  The preliminary
analysis  of  2000  run  was  presented  in  Quark  Matter   2002
\cite{na50a}.  The suppression obtained is still anomalous in the
sense that the normal nuclear absorption model fails  to  explain
it.  Compared  to  1998  run \cite{na50b}, 2000 data are flatter,
suppression being more at low and intermediate $E_T$. Preliminary
analysis indicated less suppression (compared  to  1998  run)  at
large  $E_T$, but in the final analysis suppression is compatible
with the 1998 run. 1998 NA50 data gave the  first  indication  of
anomalous  $J/\psi$ suppression and were analyzed in a variety of
models,   with   and    without    the    assumption    of    QGP
\cite{bl00,ch01,ca00,ch02,ch02a,qiu98}.  We have shown that a QCD
based nuclear absorption model, where $J/\psi$'s are absorbed  in
nuclear medium, could explain the data \cite{ch02,ch02a}. We have
also  shown  that  the  model  reproduces  the  NA50  data on the
centrality  dependence  of   $p_T$   broadening   of   $J/\psi$'s
\cite{ch03}.  What  is  more  intriguing  is  that  the predicted
$J/\psi$ over Drell-Yan ratio or the  $p_T$  broadening  at  RHIC
energy  matches  with  the  prediction  obtained in the QGP based
threshold model \cite{ch03}. It seems that even at  RHIC  energy,
$E_T$  (centrality)  dependence  of  the  $J/\psi$ suppression or
$p_T$  broadening  may  not  distinguish  a   deconfining   phase
transition.

NA50  collaboration  also  published  the analysis of the nuclear
absorption of $J/\psi$ in high statistics 450 GeV  pA  collisions
\cite{na50-a}.  They  estimated  the  $J/\psi$ nucleon absorption
cross section ($\sigma^{J/\psi N}_{abs}$)  in  the  framework  of
Glauber   model.   High   statistics   450   GeV  pA  data  yield
$\sigma^{J/\psi N}_{abs} = 4.4 \pm 1.0$  mb  \cite{na50-a}.  They
also  estimate  a common $\sigma^{J/\psi N}_{abs}$ from latest pA
and  NA38  200  GeV/c  S+U  data   \cite{na38},   $\sigma^{J/\psi
N}_{abs}$
=4.4 $\pm$ 0.5 mb.
The  extracted  absorption cross section is much smaller than the
earlier value of 6.4 $\pm$ 0.8 mb extracted from fit  to  earlier
NA50  data  \cite{na50-b} or 7.1 $\pm$ 3.0 mb obtained from a fit
to NA38 S+U data \cite{na38}. Within  error,  the  200  AGeV  S+U
cross sections are compatible with 450 AGeV pA cross sections.

In  an  earlier publication we have analysed the preliminary NA50
data of 2000 run  \cite{ch03b}.  It  was  shown  that  QCD  based
nuclear  absorption  model,  with  parameters fixed from the high
statistics  pA  data  give  consistent  description  of  $J/\psi$
suppression  in  158  AGeV Pb+Pb collisions. The preliminary data
were also analysed in the QGP based threshold model \cite{ch03a}.
In the threshold model \cite{bl00}, in addition to 'conventional'
(Glauber) nuclear absorption, an anomalous  suppression  is  used
such  that  all  the  $J/\psi$'s  are  totally suppressed above a
critical (threshold) density $n_c$. 1998  version  of  NA50  data
\cite{na50b}  were  well  explained  in the threshold model, with
$n_c$= 3.7-3.75 $fm^{-2}$ and $J/\psi$-nucleon  absorption  cross
section    $\sigma^{J/\psi    N}_{abs}$=6.4    mb    \cite{bl00}.
$J/\psi$-nucleon  absorption  cross  scetion  6.4  mb  is   large
compared  to  the  recently  extracted  value  of  from  the high
statistics   pA   data,    $\sigma^{J/\psi    N}_{abs}$=4.4    mb
\cite{na50-a}.   It   was   shown   that   with   $\sigma^{J/\psi
N}_{abs}$=4.4 mb,  the  threshold  model  fails  to  explain  the
preliminary  NA50  data \cite{na50a} on the centrality dependence
of $J/\psi$ over Drell-Yan ratio, unless the threshold density is
largely smeared \cite{ch03a}.

In  the  present  paper,  we  have  analysed the latest NA50 data
\cite{na50a1} on $J/\psi$ suppression in Pb+Pb collisions  .  The
QCD  based  nuclear  absorption model, with parameters fixed from
the high statistics pA data, still  give  consistent  description
the latest data. QGP based threshold model also explains the data
if  the  threshold  density  is  smeared.  In  addition,  we have
analysed the NA50 data on the centrality dependence  of  $J/\psi$
over  minimum  bias  ratio  \cite{na50c}  and  on  the centrality
dependence of $p_T$ broadening of $J/\psi$'s  \cite{na50d}.  Both
the  models, QCD based nuclear absorption model and the QGP based
threshold model, explain these data. The NA50 data  of  158  AGeV
Pb+Pb  collisions  could not discriminate between the two models.
It is also shown that even at RHIC energy, centrality  dependence
of $J/\psi$ suppression or $p_T$ broadening could not distinguish
between the two models.

The  paper is organised as follows: in section II we have briefly
described  the  QCD  based  nuclear  absorption  model  and   the
threshold  model.  In  section  III,  NA50 data on the centrality
dependence of $J/\psi$ over Drell-Yan ratio and the $J/\psi$ over
minimum bias ratio are analyzed. In section IV, it is shown  that
both  the  models  could  explain the NA50 $p_T$ broadening data.
Predicted centrality dependence of $J/\psi$ over Drell-Yan  ratio
and  $p_T$  broadening  at  RHIC  energy  are given in section V.
Lastly the summary and conclusions are given in section VI.

\section{Models for $J/\psi$ suppression}
\subsection{ QCD based nuclear
absorption model}

In  the  QCD  based  nuclear  absorption model \cite{ch02,qiu98},
$J/\psi$ production is assumed to be  a  two  step  process,  (a)
formation of a $c\bar{c}$ pair, which is accurately calculable in
QCD  and  (b)  formation  of a $J/\psi$ meson from the $c\bar{c}$
pair, which is conveniently  parameterized.  The  $J/\psi$  cross
section  in  $AB$ collisions, at center of mass energy $\sqrt{s}$
is written as,

\begin{eqnarray} \label{1a}
\sigma^{J/\psi} (s) &&
=K \sum_{a,b} \int dq^2 \left( \frac{\hat \sigma_{ab \rightarrow
cc}}     {Q^2}     \right)
 \int    dx_F    \phi_{a/A}(x_a,Q^2) \\ \nonumber
&& \phi_{b/B}(x_b,Q^2) \frac{x_a x_b}{x_a + x_b}
\times  F_{c\bar{c}
\rightarrow J/\psi} (q^2), \end{eqnarray}

\noindent  where  $\sum_{a,b}$  runs over all parton flavors, and
$Q^2 = q^2 +4 m_c^2$. The  $K$  factor  takes  into  account  the
higher  order corrections. The incoming parton momentum fractions
are fixed by kinematics and are $x_a
=(\sqrt{x^2_F+4Q^2/s}+x_F)/2$               and              $x_b
=(\sqrt{x^2_F+4Q^2/s}-x_F)/2$.
$\hat  \sigma_{ab \rightarrow c\bar{c}}$ are the subprocess cross
section and are given in \cite{be94}. $F_{c  \bar{c}  \rightarrow
J/\psi}(q^2)$  is  the  transition  probability that a $c\bar{c}$
pair with relative momentum square $q^2$ evolve into  a  physical
$J/\psi$ meson. It is parameterized as,

\begin{eqnarray} \label{4} F_{c \bar{c} \rightarrow J/\psi} (q^2)
= && N_{J/\psi} \theta(q^2) \theta({4m^\prime}^2 - 4 m_c^2 -q^2) \\ \nonumber
&& (1   -   \frac{q^2}{{4m^\prime}^2   -   4   m_c^2  })^{\alpha_F}.
\end{eqnarray}

In  a  nucleon-nucleus/nucleus-nucleus  collision,  the  produced
$c\bar{c}$ pairs interact with nuclear medium before  they  exit.
It  is  argued  \cite{qiu98} that the interaction of a $c\bar{c}$
pair  with  nuclear  environment  increases  the  square  of  the
relative  momentum between the $c\bar{c}$ pair. As a result, some
of the $c\bar{c}$ pairs can gain enough relative square  momentum
to   cross   the   threshold  to  become  an  open  charm  meson.
Consequently,  the  cross  section  for  $J/\psi$  production  is
reduced  in comparison with nucleon-nucleon cross section. If the
$J/\psi$ meson travel a distance $L$,  $q^2$  in  the  transition
probability  is  replaced  to $q^2 \rightarrow q^2 +\varepsilon^2
L$, $\varepsilon^2$ being the relative square momentum  gain  per
unit   length.   In   \cite{ch02},   parameters   of   the  model
($\alpha_F$,$KN_{J/\psi}$ and $\varepsilon^2$)  were  fixed  from
experimental  data  on  total  $J/\psi$  cross  section  in pA/AA
collisions, $KN_{J/\psi}$=0.458,  $\varepsilon^2=0.225  GeV^2/fm$
and $\alpha_F =1.0$ \cite{ch02}.

\begin{figure}[h]
\centerline{\psfig{figure=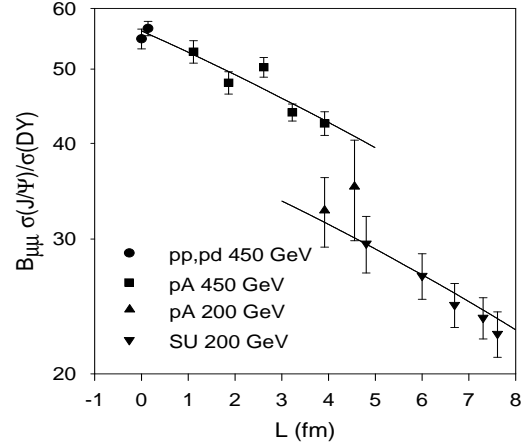,height=10cm,width=8cm}}
\vspace{-3.0cm}
\caption{  The  latest  NA50  experimental  ratio  of  the  total
$J/\psi$ cross sections and Drell-Yan cross sections in pp and pA
collisions {\protect \cite{na50-a}}. The solid lines are the  fit
obtained in the QCD based nuclear absorption model.} \end{figure}

As  mentioned  in  the  beginning,  recently  NA50  collaboration
measured $J/\psi$ cross section in pA  collisions  \cite{na50-a}.
They  have  measured  $B_{\mu\mu}  \sigma(J/\psi)/\sigma(DY)$. In
Fig.1, experimental data are shown as a function of  the  nuclear
length  $L$.  The  Drell-Yan  cross  sections donot have any A or
alternately any L-dependence. The observed $L$ dependence is then
due to $J/\psi$'s only. We fit  the  data  with  two  parameters,
$N_{norm}=KN_{J\/psi}/\sigma^{DY}_{NN}(nb)$  and  square momentum
gain factor $\varepsilon^2$ ($\alpha_F$ being kept fixed  at  1).
In Fig.1, the fit obtained to the data are shown. The two sets of
data  at  200  GeV/c  and 450 GeV/c could be fitted with a common
square momentum gain factor, $\varepsilon^2$= 0.187 $GeV^2/fm$, a
value 20\% lower than the  value  obtained  earlier  \cite{ch02}.
Lowering   of   $\varepsilon^2$   indicate   less  absorption  of
$J/\psi$'s in nuclear medium, in agreement with the Glauber model
calculations. While the square momentum gain factor do  not  show
energy  dependence,  the  evident  energy dependence of the cross
section ratios shows up in  the  other  parameter  of  the  model
$N_{norm}$.  We  obtain  $N_{norm}$  =  10.18  at  200  GeV/c and
$N_{norm}$ = 4.43 at 450 GeV/c. The energy dependence of $J/\psi$
cross section being taken care of in the model (Eq.\ref{1a}), the
energy dependence of $N_{norm}$ is due  to  the  Drell-Yan  cross
sections only. In the mass range, $2.9 > M >4.5$ GeV, the Craigie
parameterization  \cite{craigie}, of the Drell-Yan cross section,
$\sigma(DY) \propto e^{-14.9M/\sqrt{s}}$,  gives  for  the  ratio
$\sigma(DY)_{450GeV}/\sigma(DY)_{200GeV}$  = 2.1 -3.1, consistent
with the presently obtained ratio of 2.29.

With  the  parameters  of  the model fixed from high statistic pA
data, $J/\psi$ production cross section in Pb+Pb  collisions  are
obtained following the standard procedure. At an impact parameter
${\bf  b}$,  $J/\psi$  production cross section, as a function of
$E_T$ is written as \cite{ch02},

\begin{equation}
\frac{d^3\sigma^{J/\psi}}{dE_Td^2b} = \sigma^{J/\psi}_{NN}
\int d^2s T_A({\bf s}) T_B({\bf b-s}) S(L({\bf b,s})) P(b,E_T)
\end{equation}

\noindent  where  $T_{A,B}$  is  the  nuclear thickness function,
$T(b)=\int_{-\infty}^{\infty} dz \rho(b,z)$.  $P(b,E_T)$  is  the
$E_T-b$  correlation function. We have used the Gaussian form for
the $E_T-b$ correlation,

\begin{equation}\label{5}
P(b,E_T) \propto exp(-(E_T-qN_p(b))^2/2q^2aN_p(b))
\end{equation}

\noindent where $N_p(b)$ is the number of participant nucleons at
impact  parameter  b.  $a$  and  $q$  are  parameters  related to
dispersion and average transverse energy.  For  Pb+Pb  collisions
the  parameters  are,  $a$=1.27  and  $q$=0.274  GeV \cite{bl00}.
$S(L)$ is the suppression factor due to passage through a  length
L in nuclear environment. At an impact parameter ${\bf b}$ and at
point ${\bf s}$, the transverse density is calculated as,

\begin{equation} \label{den}
n({\bf b,s})=T_A({\bf s})[1-e^{-\sigma_{NN}T_B({\bf b-s})}] +
[A \leftrightarrow B]
\end{equation}

\noindent  and  the length $L({\bf b,s})$ that the $J/\psi$ meson
will traverse is obtained as,

\begin{equation}
L({\bf b,s})=n({\bf b,s})/2\rho_0
\end{equation}

At  a fixed impact parameter $E_T$ fluctuates. $E_T$ fluctuations
at a fixed impact parameter plays an important role  in  $J/\psi$
suppression.  The  second  drop  in  the  $J/\psi$ over Drell-Yan
ratio,  beyond  100 GeV is due to $E_T$ fluctuations \cite{bl00}.
Following  Blaizot  et al \cite{bl00}, we take into account $E_T$
fluctuations at a fixed impact parameter $b$, by the replacement:

\begin{equation}
L({\bf b,s}) \rightarrow L({\bf b,s})E_T/<E_T>(b).
\end{equation}

\subsection{QGP based threshold model}

We   have   also   analyzed  the  data  in  the  threshold  model
\cite{bl00}.  The  details  of  the  model  could  be  found   in
\cite{bl00}.  Briefly,  in  the  threshold  model, in addition to
Glauber  type  'nuclear'  absorption,  an  anomalous  suppression
factor  $S_{anom}$  is  used.  The  $J/\psi$  cross section at an
impact parameter ${\bf b}$ as a function of $E_T$ is then written
as,

\begin{eqnarray} \label{blaizot}
\frac{d^3\sigma^{J/\psi}}{dE_Td^2b} = &&\sigma^{J/\psi}_{NN}
\int d^2s T^{eff}_A({\bf s}) T^{eff}_B({\bf b-s})\\
&& S_{anom}({\bf b,s}) P(b,E_T) \nonumber
\end{eqnarray}

\noindent   where   $T^{eff}(b)$   is  the  effective  thickness,
$T^{eff}({\bf  b})=\int_{-\infty}^{\infty}  dz  \rho({\bf   b},z)
exp(-\sigma^{J/\psi  N}_{abs}  \int_z^{\infty} dz\prime \rho({\bf
b},z\prime)$, $\sigma^{J/\psi N}_{abs}$ is  the  $J/\psi$-nucleon
absorption  cross-section.  In  \cite{bl00},  Blaizot  et al used
$\sigma^{J/\psi N}_{abs}$=6.4 mb. In  the  present  analysis,  we
have used $\sigma^{J/\psi N}_{abs}$=4.4 mb, as extracted from the
recent pA data \cite{na50-a}. In Eq.\ref{blaizot}, $S_{anom}({\bf
b,s})$  is  the  anomalous  suppression  factor.  Blaizot  et  al
\cite{bl00} considered two types of form for $S_{anom}$. Assuming
that all the $J/\psi$'s get suppressed above a threshold  density
($n_c$), the anomalous suppression factor was written as,

\begin{equation} \label{s1}
S_{anom}({\bf b,s}) =\Theta (n({\bf b,s})-n_c)
\end{equation}

\noindent  where $n$ is the transverse density (Eq.\ref{den}). In
ref.\cite{bl00} it was seen that if the theta function is smeared
at the expense of another parameter,  such  that  suppression  is
gradual  rather  than abrupt, the quality of fit to data improves
considerably. This was implemented by writing,

\begin{equation} \label{s2}
S_{anom}({\bf b,s}) = 0.5 [1- tanh(\lambda (n({\bf b,s})-n_c))]
\end{equation}

In  both the form, effect of $E_T$ fluctuations at a fixed impact
parameter was taken into account by rescaling the density as,  $n
\rightarrow  n  E_T/<E_T>({\bf  b})$.  The  parameters  $n_c$ and
$\lambda$ are then obtained by fitting the latest  NA50  data  on
centrality dependence of $J/\psi$ over Drell-Yan ratio.

The  Drell-Yan  pairs  do not suffer final state interactions and
the cross section at an impact parameter ${\bf b}$ as a  function
of $E_T$ could be written as,

\begin{equation}
\frac{d^3\sigma^{DY}}{dE_Td^2b} = \sigma^{DY}_{NN}
\int d^2s T_A({\bf s}) T_B({\bf b-s}) P(b,E_T)
\end{equation}

\section{results}
\subsection{$E_T$ dependence of $J/\psi$ over Drell-Yan
ratio}

In Fig.2, centrality dependence of $J/\psi$ over Drell-Yan ratio,
as  obtained by the NA50 collaboration in their final analysis of
the 2000 Pb+Pb run, is shown. Just to show, how the  Pb+Pb  data
are  changed  with  time,  we  have  also  shown  the  results of
1996-1998 run and the preliminary analysis of the 2000 run. Final
analysis of 2000 run differ considerably from the  first  version
of  the  data,  presumably  due  to different analysis method. In
Fig.2,  model  predictions  for  $J/\psi$  suppression  are  also
depicted.                We               have               used
$B_{\mu\mu}\sigma^{J/\psi}_{NN}/\sigma^{DY}_{NN}=38$, 17\%  lower
than  the  value  obtain  from  extrapolating 200 AGeV pA/SU data
(Fig.1)  to pp collisions. 200 AGeV pA/SU data is limited to $L>3
fm$ and extrapolation to pp data may not be very accurate.

Just  to  show  that  the latest data are also anomalous, we have
shown  the  Glauber  model   calculation   with   $\sigma^{J/\psi
N}_{abs}$=4.4   mb   (the   dash-dot-dot  line).  Only  for  very
peripheral collisions, the Glauber model  of  nuclear  absorption
fits the data. For more central collisions, it produces much less
suppression  than  the data exhibit.   In  Fig.2,  the  solid
line is the calculated ratio in the QCD based nuclear absorption mdoel. 
It agrees well with the experiment.
The  parameters  of the model were obtained from fitting pA data.
In pA collisions we donot expect a deconfining phase  transition.
Ability  of  the  model  to  reproduce  Pb+Pb data, with the same
parameters,  clearly  indicate  that  nuclear  absorption  alone,
treated   in  an  {\em  unconventional}  manner,  is  capable  of
explaining the data.

\begin{figure}[h]
\centerline{\psfig{figure=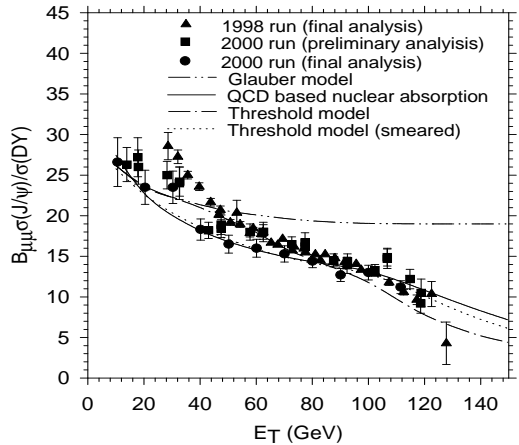,height=10cm,width=8cm}}
\vspace{-3.0cm}
\caption{The  experimental centrality dependence of $J/\psi$ over
Drell-Yan ratio in 158 AGeV Pb+Pb collisions. The  black  circles
are  the  final analysis of 2000 run. The preliminary analysis of
the 2000 run is shown as black squares and the 1st version of the
data obtained in 1996-1998  run  is  shown  as  black  triangles.
Different  lines in the figure, are: (i) Glauber model of nuclear
absorption with $\sigma^{J/\psi  N}_{abs}$=4.4  mb  (dash-dot-dot
line),  (ii) model prediction in the QCD based nuclear absorption
model (solid line), (iii)  one  parameter  threshold  model  with
threshold  density  $n_c$=3.78 $fm^{-2}$ (dash-dot line) and (iv)
two  parameter  threshold  model  calculation   with   $n_c$=3.98
$fm^{-2}$ and $\lambda$=.82 $fm^{2}$ (dotted line).} \end{figure}

The   QGP   based  threshold  model  \cite{bl00}  with  only  one
parameter, the threshold density, on the other hand fails to give
proper description of the data. In Fig.2 the dash-dot-dot line is
the  best  fit  obtained  to  the  data  with  threshold  density
$n_c$=3.78   $fm^{2}$.   In  the  intermediate  range  of  $E_T$,
agreement with data is not good.  Much  better  fit  to  data  is
obtained,  when  the  threshold  density is smeared. The dash-dot
line is the  best  fit  obtained  to  the  data  with  $n_c$=3.98
$fm^{-2}$  and $\lambda$=0.82 $fm^{2}$. The model then reproduces
the data through out the $E_T$ range. Small  value  of  $\lambda$
required for good fit to data indicate that considerable smearing
of  the  threshold  density is required for proper description of
the data. The anomalous suppression is not abrupt  but  increases
gradually  with  density.  The  threshold density $n_c$ we obtain
from fitting is also larger than the value of  3.7-3.75  obtained
by  Blaizot  et al \cite{bl00}. This is presumably due to smaller
value of the $J/\psi$-nucleon absorption cross-section.

\subsection{$E_T$ dependence of minimum bias cross section}

In   1996  and  1998  runs,  NA50  collaboration  obtained  $E_T$
dependence of the  $J/\psi$  over  minimum  bias  cross  sections
\cite{na50c}.  The  minimum  bias  cross  sections  are  easy  to
calculate. It is essentially the inelastic cross-section. In  the
Glauber  model,  at  impact parameter ${\bf b}$ and at transverse
energy $E_T$, the minimum bias cross section is written as,

\begin{equation} \label{12}
\frac{d^3\sigma^{MB}}{dE_T d^2b}  \propto (1 - exp( -\sigma_{NN} T_{AB}(b))
P(b,E_T)
\end{equation}

\noindent   where  $T_{AB}(b)=\int  d^2s  T_A({\bf  s})  T_B({\bf
s}-{\bf b})$.

\begin{figure}[h]
\centerline{\psfig{figure=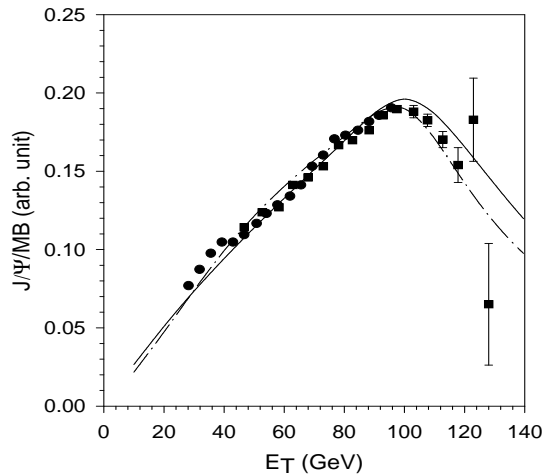,height=10cm,width=8cm}}
\vspace{-3.0cm}
\caption{ The experimental centrality dependence of $J/\psi$ over
minimum  bias ratio in 158 AGeV Pb+Pb collisions . The solid line
is the ratio obtained in the QCD based nuclear absorption  model.
The  dash-dotted  line  is  the  threshold model calculation with
$n_c$=3.98 $fm^{-2}$ and $\lambda$=.82 $fm^{2}$.} \end{figure}

In Fig.3, the NA50 data \cite{na50c} are shown. Data are given in
arbitrary  units.  The  QCD  based  nuclear absorption model with
parameters fixed from pA/AA total  $J/\psi$  cross  section  (the
solid  line)  well  describe  the  data.  Only at large $E_T$ the
predictions differ from the experiment. Data at very large  $E_T$
may  not  be  reliable.  Final  analysis  of  2000  run show less
suppression of $J/\psi$ than found in the analysis  of  1996-1998
run  (see  Fig.2).  The  $J/\psi$ over minimum bias ratio is also
well explained in the two parameter threshold model  (the  dashed
line).  Like the $J/\psi$ over Drell-Yan ratio, the $J/\psi$ over
minimum bias ratio is also incapable  of  distinguishing  between
the two models.

\section{$p_T$ broadening of $J/\psi$}

It  is  well  known  that  in pA and AA collisions, the secondary
hadrons generally  shows  a  $p_T$  broadening  \cite{ptbr,kh97}.
Kharzeev  et al \cite{kh97} suggested $p_T$ broadening as a probe
for the deconfining transition. They argued that in a deconfining
medium, $p_T$ broadening will visibly decrease at large $E_T$, in
contrast to a gradually increasing $p_T$ broadening in a  nuclear
medium.  Recently  we have shown that decreasing $p_T$ broadening
can  not  be  considered  as  a  signal  of   deconfining   phase
transition,  as  such  a  trend is also obtained in the QCD based
nuclear absorption model \cite{ch03}.  It  was  also  shown  that
$E_T$   fluctuations,  at  a  fixed  impact  parameter  plays  an
important role in  explaining  the  experimental  data  on  $p_T$
broadening  of $J/\psi$'s. The QCD based nuclear absorption model
\cite{ch02} could explain the NA50 $p_T$ broadening data  if  the
effect  of  $E_T$  fluctuations  are  properly  accounted for. In
\cite{ch03}, we  have  also  obtained  $p_T$  broadening  in  the
threshold  model, with a single parameter, the threshold density,
$n_c$=3.7 $fm^{-2}$ and  $\sigma^{J/\psi  N}_{abs}$=6.4  mb.  One
parameter threshold model could not fit the data. As it was shown
in    last    two   sections,   with   the   latest   NA50   data
\cite{na50a,na50-a}, parameters of both the  models  are  changed
and it interesting to see the consequence on $p_T$ broadening.

The  natural  basis for the $p_T$ broadening is the initial state
parton  scatterings.  For  $J/\psi$'s,  gluon  fusion  being  the
dominant  mechanism  for  $c\bar{c}$  production,  initial  state
scattering   of   the   projectile/target   gluons    with    the
target/projectile   nucleons   causes   the   intrinsic  momentum
broadening of  the  gluons,  which  is  reflected  in  the  $p_T$
distribution  of  the  resulting  $J/\psi$'s.  Parameterizing the
intrinsic transverse momentum of a gluon, inside a nucleon as,

\begin{equation}
f(q_T) \sim exp(-q^2_T/<q^2_T>)
\end{equation}

\noindent  momentum  distribution of the resulting $J/\psi$ in NN
collision is obtained by convoluting two  such  distributions,

\begin{equation}
f^{J/\psi}_{NN}(p_T) \sim exp(-p^2_T/<p^2_T>^{J/\psi}_{NN})
\end{equation}

\noindent  where  $<p^2_T>^{J/\psi}_{NN}  =  <q^2_T>+<q^2_T>$. In
nucleus-nucleus collisions at  impact  parameter  ${\bf  b}$,  if
before  fusion, a gluon undergo random walk and suffer $N$ number
of subcollisions, its square momentum  will  increase  to  $q^2_T
\rightarrow  q^2_T  +  N\delta_0$,  $\delta_0$  being the average
broadening in each subcollisions.  Square  momentum  of  $J/\psi$
then easily obtained as,

\begin{equation} \label{1}
<p^2_T>^{J/\psi}_{AB}(b) = <p^2_T>^{J/\psi}_{NN} + \delta_0 N_{AB}({\bf b})
\end{equation}

\noindent  where $N_{AB}({\bf b})$ is the number of subcollisions
suffered by the projectile and target gluons with the target  and
projectile nucleons respectively.

Average number of collisions $N_{AB}({\bf b})$ can be obtained in
a  Glauber  model \cite{kh97}. At impact parameter ${\bf b}$, the
positions $({\bf s},z)$ and $({\bf b-s},z^\prime)$ specifies  the
formation  point  of $c\bar{c}$ in the two nuclei, with ${\bf s}$
in the transverse plane and $z,z^\prime$ along the beam axis. The
number of collisions, prior to $c\bar{c}$ pair formation, can  be
written as,

\begin{eqnarray} \label{2}
N(b,s,z,z^\prime)    = &&   \sigma_{gN}    \int_{-\infty}^z    dz_A
\rho_A(s,z_A) \\ \nonumber
&& + \sigma_{gN} \int_{-\infty}^{z^\prime} dz_B
\rho_B(b-s,z^\prime)
\end{eqnarray}

\noindent where $\sigma_{gN}$ is the gluon-nucleon cross section.
Above  expression  should  be  averaged  over  all  positions  of
$c\bar{c}$ formation with  a  weight  given  by  the  product  of
nuclear densities and survival probabilities $S$,

\begin{eqnarray}\label{3}
&&N_{AB}(b)=
\int d^2 s \int^\infty_{-\infty} dz \rho_A(s,z)
\int^\infty_{-\infty}
dz^\prime \rho_B(b-s,z^\prime) \times  \\  \nonumber
&& S(b,s,z,z^\prime) N(b,s,z,z^\prime) /
\int d^2s
\int^\infty_{-\infty} dz \rho_A(s,z) \times \\ \nonumber
&& \int^\infty_{-\infty}
dz^\prime \rho_B(b-s,z^\prime) S(b,s,z,z^\prime)
\end{eqnarray}

Finally,  corresponding quantity at fixed transverse energy $E_T$
is obtained as,

\begin{eqnarray}\label{4a}
N_{AB}(E_T) = &&\int d^2 b P(b,E_T) \sigma_{AB} N_{AB}(b) / \\ \nonumber
&& \int d^2b P(b,E_T) \sigma_{AB}
\end{eqnarray}

\noindent  where $\sigma_{AB}$ is the inelastic cross section for
AB collisions.

\begin{figure}[h]
\centerline{\psfig{figure=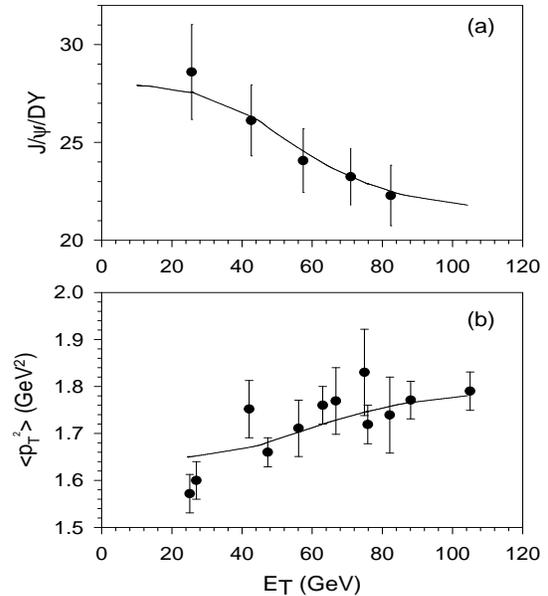,height=10cm,width=8cm}}
\vspace{-0.5cm}
\caption{ (a) NA38 experimental data on the centrality dependence
of  $J/\psi$  over  Drell-Yan ratio, in 200 GeV/c S+U collisions.
The solid line is the ratio obtained in  the  QCD  based  nuclear
absorption   model.   (b)  The  centrality  dependence  of  $p_T$
broadening in S+U collisions. The solid line is a fit to the data
in the QCD based nuclear absorption model.} \end{figure}

Fluctuations  of  $E_T$  at  a  fixed  impact parameter will also
affect the average number of collisions  $N_{AB}(E_T)$.  We  have
taken into account the $E_T$ fluctuations by the replacement,

\begin{equation}
N_{AB}(b) \rightarrow E_T/<E_T>(b) N_{AB}(b).
\end{equation}

\begin{figure}[h]
\centerline{\psfig{figure=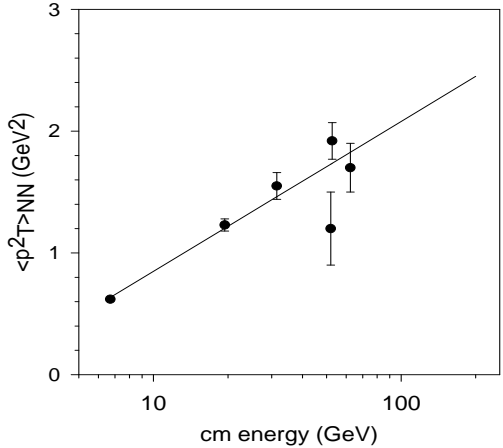,height=10cm,width=8cm}}
\vspace{-3.0cm}
\caption{  The energy dependence of $<p^2_T>_{NN}$ along with the
fit to it with Eq.\ref{6}.} \end{figure}

$p_T$  broadening  of  $J/\psi$'s in AA collisions depends on two
parameters,  (i)  $<p^2_T>_{NN}$,  the  mean  squared  transverse
momentum  in  NN  collisions,  a measurable quantity and (ii) the
product of the gluon-nucleon cross section and the average parton
momentum broadening per collision,  $\sigma_{gN}\delta_0$.  Since
gluons  are  not  free,  the  second  quantity is essentially non
measurable. We obtain $\sigma_{gN}\delta_0$ from  a  fit  to  the
NA38  $p_T$  broadening data \cite{na38} in S+U collisions at 200
GeV/c. $<p^2_T>_{NN}$ at corresponding energy is known  from  NA3
experiment,  $<p^2_T>_{NN}  =  1.23  \pm  0.05$  \cite{na3}.  The
$E_T-b$ correlation parameters, $a$ and $q$  for  S+U  collisions
are,  $a$=3.2  and  $q$=0.74  GeV  \cite{vo99}.  To show that the
present 'unconventional' nuclear absorption model also reproduces
the centrality dependence of $J/\psi$ over Drell-Yan ratio in S+U
collisions, in Fig.4a, we have  compared  our  results  with  the
experimental   data.  We  have  neglected  the  effect  of  $E_T$
fluctuations in S+U collisions. The agreement  between  data  and
theory  is  good.  In Fig.4b, NA38 experimental data on the $E_T$
dependence of $p_T$ broadening are shown. The solid line is a fit
to the data,  obtained  with  $<p^2_T>_{NN}$=  1.23  (fixed)  and
$\sigma_{gN}\delta_0$    =    $0.442   \pm   0.056$.   Value   of
$\sigma_{gN}\delta_0$ agrees closely with the value  obtained  by
Kharzeev et al \cite{kh97} in the conventional nuclear absorption
model  and  also  with  the  value  obtained in the comover model
\cite{ar99}.

$<p^2_T>_{NN}$   increases   weakly   with   energy.   To  obtain
$<p^2_T>_{NN}$ for Pb+Pb collisions at 158 GeV/c, we have  fitted
the  existing  experimental  data \cite{na3,ba78,nag75,cl78} with
logarithmic energy dependence,

\begin{equation}  \label{6}  <p^2_T>_{NN}  =  a  + b \ln \sqrt{s}
\end{equation}

In  Fig.5, experimental data along with the fitted curve obtained
with  $a=-0.38$  and  $b=0.53$.  is   shown.   From   the   above
parameterization, we obtain, $<p^2_T>_{NN}$
=1.15 $GeV^2$, for Pb+Pb collisions at CERN SPS. As we intend  to
predict  $p_T$  broadening at RHIC energy, $<p^2_T>_{NN}$ at RHIC
energy ($\sqrt{s}$=200 GeV)  is  also  obtained  from  the  above
parameterization.  At  RHIC energy, $<p^2_T>_{NN}$ =2.45 $GeV^2$.
However, we must warn our reader to treat the above  number  with
caution.  The  experimental  data  being  limited to 60 GeV only,
extrapolation to RHIC energy is unreliable.

\begin{figure}[h]
\centerline{\psfig{figure=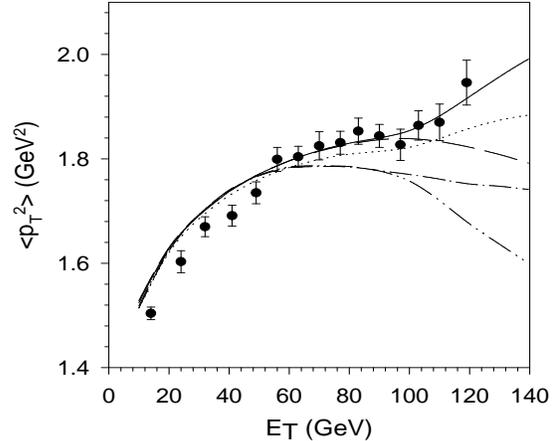,height=10cm,width=8cm}}
\vspace{-3.0cm}
\caption{   The   experimental  centrality  dependence  of  $p_T$
broadening in Pb+Pb collisions. The solid and  dashed  lines  are
the  QCD  based  nuclear  absorption  model predictions, with and
without the effect of $E_T$ fluctuations on the average number of
gluon-nucleon  collisions  $N_{AB}(b)$.  The   dash-dot-dot   and
dash-dot  lines  are  the  same  in  the  QGP based one parameter
threshold model. The prediction in the  two  parameter  threshold
model is shown as dotted line.} \end{figure}

In  Fig.6,  we  have  shown the result of $p_T$ broadening in the
model. The solid and dashed lines are the $<p^2_T>$  in  our  QCD
based  nuclear  absorption  model, with and without the effect of
$E_T$ fluctuations on $N_{AB}(b)$. When  the  $E_T$  fluctuations
are  not  accounted  for  (the  dashed line), the model could not
explain the  experiment  beyond  $E_T$=100  GeV.  Experimentally,
$<p^2_T>$  continues  to  increase with $E_T$ beyond 100 GeV, but
the model predicts a decreasing trend. Beyond 100 GeV  (the  knee
of  the  $E_T$ distribution), $J/\psi$'s are strongly suppressed.
Strong suppression causes the $p_T$ broadening to decrease beyond
100 GeV. The decreasing trend is changed into an increasing trend
if the effect of $E_T$ fluctuations on  $N_{AB}$  is  taken  into
account   (the   solid   line).  $E_T$  fluctuations  effectively
increases  the  average  number  of  collisions  $N_{AB}(b)$  and
counter  balance the strong suppression effect beyond the knee of
the $E_T$ distribution. Considering that all  the  parameters  of
the model were fixed, model describe the data very well.

In  Fig.6,  the  one  parameter  threshold  model prediction with
threshold density $n_c$=3.78 $fm^{-2}$ is shown. When the  effect
of  $E_T$ fluctuations on $N_{AB}$ is neglected (the dash-dot-dot
line) the model fails to explain  the  data.  At  low  $E_T$,  it
predict   $<p^2_T>$  in  accordance  to  the  QCD  based  nuclear
absorption model, but beyond $E_T$=60 GeV, it predict less  $p_T$
broadening.  Also,  the increasing tendency beyond 100 GeV is not
reproduced. Even when the effect of $E_T$ fluctuations are  taken
into  account (the dash-dot line), the model fails to give proper
description to the data. This is expected as  the  one  parameter
threshold   model   donot  give  very  good  description  to  the
centrality dependence of $J/\psi$ over Drell-Yan  ratio.  Huefner
et al \cite{hu02} also analyzed the NA50 $p_T$ broadening data in
the  threshold model and essentially obtain a similar result that
the model could not fit the NA50 $p_T$ broadening data. As  shown
earlier,  a two parameter threshold model, with smeared threshold
density, reproduced the centrality dependence  of  $J/\psi$  over
Drell-Yan  ratio  as well as $J/\psi$ over minimum bias ratio. In
Fig.6, the prediction obtained in  the  two  parameter  threshold
model,  with  the effect of $E_T$ fluctuations included, is shown
as the dotted line. Considering that it is also a parameter  free
calculation ($<p^2_T>_{NN}$ and $\sigma_{gN}\delta_0$ fixed), the
model  describes the data well. Here, we may mention that Armesto
et al \cite{ar99}, in the comover model also explained  the  NA50
$p_T$  broadening  data. However, as they did not account for the
$E_T$ fluctuations on $N_{AB}(b)$ the increasing trend beyond 100
GeV could not be reproduced.

\begin{figure}[h]
\centerline{\psfig{figure=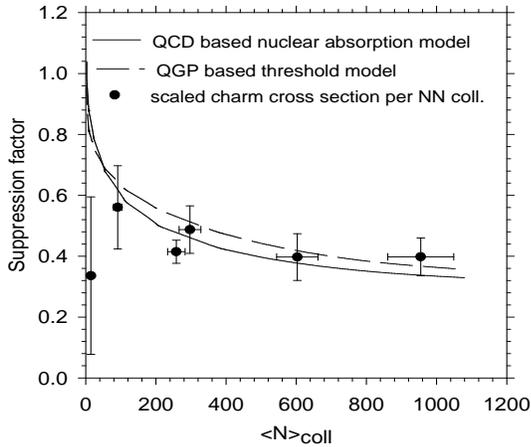,height=10cm,width=8cm}}
\vspace{-2.5cm}
\caption{   Suppression  factor  for  $J/\psi$  production  as  a
function of average number of NN collisions in Au+Au collisions 
at RHIC  energy.  The
solid  and dashed lines are the prediction from QCD based nuclear
absorption model and the QGP based threshold model  respectively.
In  the  figure,  the filled circles represent the PHENIX data on
centrality  dependence  of  total   charm   production (scaled by a 
factor of 1500)  per   NN
collisions. }
\end{figure}

\section{Prediction for RHIC Au+Au collisions}

Recently PHENIX collaboration published the centrality dependence
of   charm   production   in  Au+Au  collisions  at  RHIC  energy
\cite{phenix1} . Centrality dependence of charm quark  production
is    consistent   with   binary   collisions   scaling.   PHENIX
collaboration  also  published  the  yield  of  $J/\psi$ in a few
centrality  ranges   of   Au+Au   collisions   at   RHIC   energy
\cite{phenix2}. Data have very large error bars. Data do not show
any  indication of large enhancement as speculated in some models
\cite{th01}. We have shown  that  the PHENIX data on the centrality  
dependence  of
$J/\psi$  production  are  well  described in the QCD based
nuclear absorption model \cite{ch03e}. In this section we compare
the $J/\psi$ production at RHIC energy in the QCD  based  nuclear
absorption  model  with the production in the QGP based threshold
model. Model parameters are kept fixed at the values required for
Pb+Pb collisions at SPS energy. At RHIC energy the so called hard
scattering, proportional to number of binary  collisions  appear.
However,   PHENIX  data on $J/\psi$ production \cite{phenix2} do not
require  any
hard scattering component \cite{ch03e} and presently we neglect them.
 
In  Fig.7,  we  have compared the suppression factor for $J/\psi$
production  at  RHIC  energy  in the QCD based nuclear absorption
model and in the QGP based threshold model. In a centrality range
of collisions, $J/\psi$ suppression factor is defined as,

\begin{equation}
S=\frac{\sigma^{J/\psi}_{AA}}{<N_{AA}>\sigma^{J/\psi}_{NN}}
\end{equation}

\noindent where $<N>_{AA}$ is the average number of NN collisions
in the  centrality range of collisions. As seen in Fig.7, both the
QCD based nuclear absorption model and the  QGP  based  threshold
model predict nearly same suppression factor.
In  FIg.7  we  have  also  shown the recent PHENIX measurement of
total charm quark  multiplicity  per  NN  collisions ($N_{c\bar{c}}/T_{AB}$) \cite{phenix1}.  
Data  were
scaled  by  a  factor  of  1500.  Except  for the very peripheral
collisions (60-92\% centrality), centrality dependence  of  charm
production  agree  well with the dependence predicted in both the
models. Possibly centrality  dependence  of  $J/\psi$  production
will not distinguish between the models.

As  told earlier, $p_T$ broadening of $J/\psi$ is considered as a
probe of deconfinement transition. We have seen that at SPS $p_T$
broadening do not distinguish  between  the  QCD  based  nuclear
absorption  model  and QGP based threshold model. In Fig.8, model
predictions for $p_T$ broadening of $J/\psi$'s at RHIC energy are
shown.  Both the models predict very similar $p_T$ broadening. As
it is in SPS energy, at RHIC  energy  also  $p_T$  broadening  of
$J/\psi$  do  not  distinguish  between  the  QCD  based  nuclear
absorption and QGP bases threshold models.

\begin{figure}[h]
\centerline{\psfig{figure=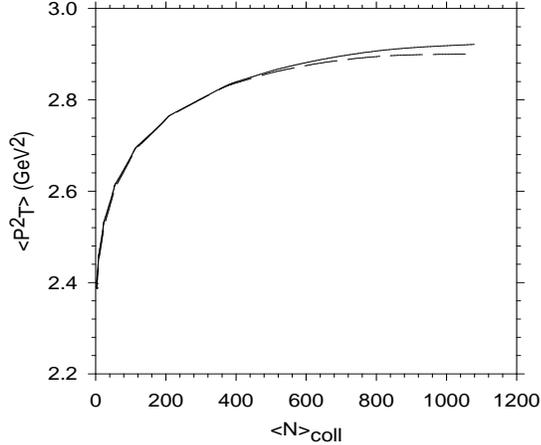,height=10cm,width=8cm}}
\vspace{-02.5cm}
\caption{ $p_T$ broadening of $J/\psi$'s as a function of average
number  of  NN  collisions  at  RHIC energy. The solid and dashed
lines are the prediction from QCD based nuclear absorption  model
and the QGP based threshold model respectively.} \end{figure}

\section{summary and conclusions}

To summarize, we have analyzed the latest (2000 run) NA50 data on
$J/\psi$  suppression  in  158  AGeV  Pb+Pb collisions. QCD based
nuclear absorption model, with parameters  fixed  from  the  NA50
high  statistics  pA data, well explain the latest NA50 data. The
model also explain the centrality  dependence  of  $J/\psi$  over
minimum  bias ratio as well as the centrality dependence of $p_T$
broadening of $J/\psi$'s. The same data sets were analysed in the
 QGP based threshold model. Threshold model with smeared
threshold density also explain those data sets. At SPS energy $J/\psi$'s
can not distinguish between the two differeny models.
We have also shown that  even  at  RHIC
energy,   both  the  models  predict  nearly  similar  centrality
dependence of $J/\psi$  suppression  and  its  $p_T$  broadening.
Possibly,  for  the  deconfinement  phase transition, there is no
'smoking gun'. A variety of data, analysed in a consistent manner
will be able to shed light on the  possible  deconfinement  phase
transition.

\end{document}